\documentclass[prd,aps,showpacs,nofootinbib,tightenlines]{revtex4}
\usepackage{amsmath}
\usepackage{bm}
\usepackage{times}
\usepackage{braket}
\usepackage{color}
\usepackage{epsfig}
\usepackage{slashed}
\usepackage{hyperref}
\usepackage{amssymb}
\usepackage{graphicx}
\newcommand{\beq}{\begin{eqnarray}}
\newcommand{\eeq}{\end{eqnarray}}
\newcommand{\non}{\nonumber\\ }

\def \cpc{ Chin. Phys. C  }

\def \epja{ Eur. Phys. J. A }
\def \epjc{ Eur. Phys. J. C }
\def \ijmpa{ Int. J. Mod. Phys. A }

\def \npb{  Nucl. Phys. B }
\def \plb{  Phys. Lett. B }

\def \pr{  Phys. Rev. }
\def \prd{  Phys. Rev. D }
\def \prl{  Phys. Rev. Lett.  }

\def \zpc{  Z. Phys. C }

\definecolor{Red}{rgb}{1.,0.,0.}

\definecolor{Blue}{rgb}{0.,0.,1.}

\definecolor{nicered}{rgb}{0.7,0.1,0.1}
\definecolor{nicegreen}{rgb}{0.1,0.5,0.1}

\hypersetup{colorlinks,citecolor=nicegreen,linkcolor=nicered}
\begin{document}
\title{Quasi-two-body decays $B_{(s)} \to  D (\rho(1450),\rho(1700)) \to D \pi \pi$ in the perturbative QCD factorization approach}
\author{Ai-Jun Ma$^1$}  \email{theoma@163.com}
\author{Ya Li$^1$}\email{liyakelly@163.com}
\author{Wen-Fei Wang$^2$}\email{wfwang@sxu.edu.cn}
\author{Zhen-Jun Xiao$^{1,3}$}\email{xiaozhenjun@njnu.edu.cn}
%
\affiliation{$^1$ Department of Physics and Institute of Theoretical Physics,
                          Nanjing Normal University, Nanjing, Jiangsu 210023, People's Republic of China}
\affiliation{$^2$ Institute of Theoretical Physics, Shanxi University, Taiyuan, Shanxi 030006, People's Republic of China}
\affiliation{$^3$ Jiangsu Key Laboratory for Numerical Simulation of Large Scale Complex
Systems, Nanjing Normal University, Nanjing, Jiangsu 210023, People's Republic of China}
\date{\today}

\begin{abstract}
By employing a framework for the quasi-two-body decays in the perturbative QCD (PQCD) factorization approach,
we calculate the branching ratios of the decays $B_{(s)} \to  D (\rho(1450),\rho(1700))\to
D \pi \pi$ with $D=(D_{(s)}, \bar{D}_{(s)})$.
The pion vector form factor $F_\pi$, acquired from a {\it BABAR} Collaboration analysis of $e^+ e^- \to \pi^+ \pi^-(\gamma)$
data, is involved in the two-pion distribution amplitudes $\Phi^{I=1}_{\pi\pi}$.
The PQCD predictions for the branching ratios of the considered quasi-two-body decays
are in the range of $10^{-10} \sim 10^{-4}$.
The PQCD predictions for ${\cal B}(B^0\to \bar{D}^0 (\rho^0(1450),\rho^0(1700)) \to \bar{D}^0 \pi^+\pi^-)$
agree well with the measured values as reported by LHCb if one takes still large theoretical
errors into account. Unlike the traditional way of the PQCD approach,
one can extract the decay rates for the two-body decays $B_{(s)} \to  D (\rho(1450),\rho(1700))$
from  the results of the corresponding quasi-two-body decays.
The PQCD predictions for ${\cal B}(B_{(s)} \to  D \rho(1450))$ and ${\cal B}(B_{(s)} \to  D \rho(770))$
are similar in magnitude: an interesting relation to be tested by future experimental measurements.
\end{abstract}

\pacs{13.25.Hw, 12.38.Bx, 14.40.Nd}

\maketitle

\section{Introduction}\label{sec:1}

Up to now, many hadronic charmed three-body $B_{(s)}$ meson decays have been measured by experiments, such as the Belle, {\it BABAR}, D0, CDF,
and LHCb Collaborations~\cite{1612.07233,plb553-159,prd76-012006,prd79-112004,prd92-032002,prd94-072001}.
The study for the $B_{(s)} \to D h h^\prime$ decays with $D=(D_{(s)}, \bar{D}_{(s)})$ and $h^{(')}=(\pi,K)$, for example,
has played an important role in the precise determination of the Cabibbo-Kobayashi-Maskawa (CKM) angle $\gamma$ ~\cite{1702.01535} and the study of the rich resonant structure~\cite{prd92-032002}.
Recently, a study of the $\pi^+\pi^-$ system by the LHCb Collaboration~\cite{prd92-032002} was performed
through Dalitz plot analysis~\cite{pm44-1068} of $B^0 \to \bar{D}^0 \pi^+ \pi^-$ decays.
The phase-space range was broad, from $0.28$ ($\approx 2 m_\pi$) to $3.4$ GeV($\approx m_B-m_D$),
and the first observation of the decay $B^0 \to \bar{D}^0 \rho^0(1450)$ was reported \cite{prd92-032002}.
When a decay rate of $\rho^0(1450) \to \pi^+\pi^-$ determined by employing the Isobar
model~\cite{pr135-B511,pr166-1731,prd11-3165} or the $K$-matrix formalism~\cite{epja16-229} was used, respectively,
the LHCb Collaboration reported their measurements for the branching fraction of the quasi-two-body decay
$B^0 \to \bar{D}^0 \rho^0(1450)$ \cite{prd92-032002},
\begin{eqnarray}
{\cal B}(B^0 \to \bar{D}^0 \rho^0(1450) \to \bar{D}^0 \pi^+ \pi^-)&=& \left\{\begin{array}{ll}
1.36 \pm 0.28 \pm 0.08 \pm 0.19 \pm 0.06 \times 10^{-5}\; {\rm (Isobar)},\\
1.91 \pm 0.37 \pm 0.73 \pm 0.19 \pm 0.09 \times 10^{-5}\; {\rm (K-matrix)}\\
\end{array} \right.
\end{eqnarray}
Meanwhile, the branching fraction of the quasi-two-body decay
$B^0 \to \bar{D}^0 \rho^0(1700)$ with a given $\rho^0(1700) \to \pi^+\pi^-$ was also reported
in Ref.~\cite{prd92-032002},
\begin{eqnarray}
{\cal B}(B^0 \to \bar{D}^0 \rho^0(1700) \to \bar{D}^0 \pi^+ \pi^-)&=& \left\{\begin{array}{ll}
0.33 \pm 0.11 \pm 0.06 \pm 0.05 \pm 0.02 \times 10^{-5}\; {\rm (Isobar)},\\
0.73 \pm 0.18 \pm 0.53 \pm 0.10 \pm 0.03 \times 10^{-5}\; {\rm (K-matrix)}.\\
\end{array} \right.
\end{eqnarray}
Similar quasi-two-body decays like $B^0 \to K^+ \rho^-(1450) $ and $B^- \to \pi^- \rho^0(1450) $
have been observed by the {\it BABAR} Collaboration ~\cite{prd83-112010,prd79-072006} with the cascade decay
$\rho(1450) \to \pi \pi$.
For $\rho(1450)$ and $\rho(1700)$ \footnote{ In the following sections, we generally use the abbreviation $\rho=\rho(770)$,
$\rho'=\rho(1450)$, $\rho''=\rho(1700)$, $\rho'''=\rho(2254)$, and $D=(D_{(s)}, \bar{D}_{(s)})$. },
there is a strong interference near 1.6 GeV.
High-statistics study of the $\tau^- \to \pi^-\pi^0\nu_\tau$ decay by Belle~\cite{prd78-072006}
reported the first observation of both $\rho'$ and $\rho''$.
In the study of $e^+ e^- \to \pi^+ \pi^-(\gamma)$ by {\it BABAR}~\cite{prd86-032013}, a clear picture of
the two $\pi^+ \pi^-$ resonances interfering with the $\rho$ was reported.
The basic properties of $\rho'$ and $\rho''$ mesons from PDG2016 ~\cite{PDG2016} are listed in Table~\ref{tab:pro}.

On the theory side, the three-body $B_{(s)}$ decays have been investigated by employing the QCD factorization approach
~\cite{plb622-207,prd74-114009,prd79-094005, prd81-094033,
prd66-054015,prd76-094006,prd88-114014,prd89-074025,prd94-094015,prd89-094007,npb899-247,epjc75-536,prd87-076007,prd70-034033},
the perturbative QCD (PQCD) factorization approach~
\cite{plb561-258,prd70-054006,prd89-074031,plb763-29,prd91-094024,epjc76-675,ma17,prd95-056008,1701.01844,1701.02941,1704.07566},
and the framework of the symmetry principles ~\cite{prd72-075013,prd72-094031,prd84-056002,
plb726-337,prd89-074043,plb728-579,prd91-014029,prd92-054010,prd84-034040}.
In three-body $B$ decays, there are two distinct final state interaction mechanisms:
(a) the interactions between the meson pair in the resonant region associated with various intermediate states,
and (b) the rescattering between the third particle and the pair of mesons usually ignored in the quasi-two-body
approximation.
In the real data analysis, most of the quasi-two-body decays are extracted from the Dalitz-plot analysis
of the three-body ones; the study of quasi-two-body $B$ decays could be a starting point in the studies
of the three-body decays.

The two-body $B \to D \rho$ decays  have been studied intensively by using various theoretical methods or
approaches~\cite{prd60-074029,prd65-096007,prd69-094018,prd75-074021,prd78-014018}.
But the three-body $B$ decays involving the radially excited $\rho$ mesons ($\rho', \rho''...$)
have not attracted much attention in the literature.
Recently, a study of the $\pi^+\pi^-$ system was performed through Dalitz-plot analysis of
$B^0 \to \bar{D}^0 \pi^+ \pi^-$ decays by the LHCb Collaboration~\cite{prd92-032002}, while
the quasi-two-body decays $B \to K (\rho,\rho') \to K \pi \pi$ were investigated
by using the PQCD approach ~\cite{plb763-29}.
The resonant and nonresonant contributions between the $\pi\pi$ pair were parametrized
into the timelike pion form factors involved in the two-pion distribution
amplitudes~\cite{Grozin,Muller,plb561-258,prd70-054006,prl81-1782,npb555-231}.
Besides $\rho$, the contribution from the $\rho'$ intermediate state could also be singled out from the
given timelike form factor $F_\pi$.
By using the Gegenbauer moments of the $P$-wave two-pion distribution amplitudes,
we can make the predictions for the branching ratios and the direct $CP$ asymmetries of the
$B \to K \rho' \to K \pi \pi$ decays. Following Ref.~\cite{plb763-29},  we have studied the
$B_{(s)} \to P \rho (\rho',\rho'') \to P \pi \pi$~\cite{prd95-056008,1704.07566}
and $B_{(s)} \to D \rho \to D \pi \pi$ decays \cite{ma17}
where $P$ denotes the light pseudoscalar mesons $\pi, K, \eta ,\eta^\prime$ and $D$ stands for
the charmed $D$ meson.

Based on our previous works in Refs.~\cite{plb763-29,ma17,prd95-056008,1704.07566},
we here study all $B_{(s)} \to  D (\rho', \rho'') \to D \pi \pi$ decay modes
and present the PQCD predictions for their branching ratios.
The typical Feynman diagrams that may contribute to the considered
decay modes are the same ones as those illustrated in Figs.~1 and 2 of Ref.~\cite{ma17}.
Since only tree operators are involved here, the direct $CP$-violating
asymmetries for the considered decays are absent  naturally.
Without the information about the distribution amplitudes for $\rho'$ and $\rho''$, the PQCD approach
does not work in calculating the branching ratios of the two-body decays $B_{(s)} \to  D (\rho',\rho'')$ in a
traditional way.
Unlike the observed ${\cal B}(\rho\to \pi\pi) \sim 100\%$, the $\rho'$ also has other decay channels like $4\pi$, $K\bar{K}$, $\omega \pi$, etc.~\cite{PDG2016}. For $\rho''$, we also know that $\rho''\to \rho \pi\pi$ is dominant \cite{PDG2016}.
In the quasi-two-body framework, fortunately, we can extract the branching ratios for the two-body decays
$B_{(s)} \to  D (\rho',\rho'') $ from the results of $B_{(s)} \to  D (\rho',\rho'')\to  D \pi\pi $
after making a reliable estimation for the branching fraction ${\cal B}((\rho',\rho'') \to \pi\pi)$.
This paper is organized as follows. In Sec.~II, we give a brief introduction for the theoretical framework.
The numerical values, some discussions and the conclusions are given in last two sections.

\begin{table}[]\label{tab:pro}
\caption{The main properties of $\rho'$ and $\rho''$ mesons~\cite{PDG2016}.}
\begin{tabular}{lcccc}
\hline\hline
Mesons          &$I^G$   &$J^{PC}$              &Mass (MeV) & Width (MeV)
 \\ \hline
$\rho'$  & $1^+$ & $1^{--}$ & $1465\pm 25$ & $400\pm 60$\\
$\rho''$  & $1^+$ & $1^{--}$ & $1720\pm 20$ & $250\pm 100$\\
\hline\hline
\end{tabular}
\end{table}


\section{The theoretical framework}\label{sec:2}

For the considered $B_{(s)} \to  D (\rho',\rho'') \to D \pi \pi$
decays, the effective Hamiltonian is of the form
\begin{eqnarray}
{\cal  H}_{eff}&=& \left\{\begin{array}{ll}
\frac{G_F}{\sqrt{2}}V^*_{cb}V_{uq}\left[C_1(\mu)O_1(\mu)+C_2(\mu)O_2(\mu)\right],
& \ \  {\rm for} \ \ B_{(s)} \to   \bar{D}_{(s)} (\rho',\rho'') \to  \bar{D}_{(s)} \pi \pi\ \ {\rm decays},\\
\frac{G_F}{\sqrt{2}} V^*_{ub}V_{cq}\left[C_1(\mu)O_1(\mu)+C_2(\mu)O_2(\mu)\right],
& \ \  {\rm for} \ \ B_{(s)} \to  D_{(s)} (\rho',\rho'') \to D_{(s)}\pi \pi\ \ {\rm decays},\\
\end{array} \right.
\end{eqnarray}
where $O_{1,2}(\mu)$ represent the tree operators, $C_{1,2}(\mu)$ are the Wilson coefficients, $q=(d,s)$, and $V_{ij}$
are the CKM matrix elements.

In the framework of the PQCD approach for the quasi-two-body decays,
the nonperturbative dynamics associated with the pair of the pion mesons are factorized into
two-meson distribution amplitudes~\cite{Grozin,Muller,plb561-258,prd70-054006,prl81-1782,npb555-231}
due to two reasons~\cite{plb561-258,prd70-054006}.
First, it is not practical to make a direct evaluation for the hard $b$-quark decay kernels
containing two virtual gluon propagators at leading order, while the possible contribution in such a kinematic region
is also power suppressed and not important.
Secondly, the dominant contribution most possibly comes from the region where the
involved two energetic mesons are almost collimating to each other and having an invariant mass
below $O(\bar\Lambda m_B)$ ($\bar\Lambda=m_B-m_b$).

Analogous to the two-body $B$ decays, the decay amplitude $\cal A$ for the quasi-two-body decays
$B_{(s)} \to  D  (\rho',\rho'') \to D \pi \pi$
in the PQCD approach can be written conceptually as the convolution ~\cite{plb561-258,prd70-054006}
\begin{eqnarray}
{\cal A}=\Phi_B\otimes H\otimes \Phi_{D}\otimes\Phi^{I=1}_{\pi\pi},
\end{eqnarray}
where the hard kernel $H$ describes the dynamics of the strong and electroweak interactions in the decays,
$\Phi_B$, $\Phi_D$ and $\Phi_{\pi\pi}$ denote the distribution amplitudes for the $B_{(s)}$ meson,
the final state $D=(D_{(s)}, \bar{D}_{(s)})$ meson and the final state $\pi\pi$ pair.
In this work, the widely used wave functions for $B_{(s)}$ meson
and $D$ mesons as used for example in Refs.~\cite{ma17,prd95-056008} are adopted.

For the $(\rho^\prime,\rho^{\prime \prime})$ mesons, their longitudinal distribution amplitudes are defined
in the same way as in Ref.~\cite{plb763-29},
\begin{eqnarray}
\Phi^{I=1}_{\pi\pi}=\frac{1}{\sqrt{2N_c}}\left[p \hspace{-2.0truemm}/\phi^0(z,\zeta,w^2)
+w\phi^s(z,\zeta,w^2) +\frac{p \hspace{-2.0truemm}/_ 1p \hspace{-2.0truemm}/_2-p \hspace{-2.0truemm}/_2p \hspace{-2.0truemm}/_ 1}{w(2\zeta-1)} \phi^t(z,\zeta,w^2)\right]\; ,
\end{eqnarray}
with the functions
\begin{eqnarray}
\phi^0(z,\zeta,w^2)&=& \frac{3F_\pi(s)}{\sqrt{2N_c}}z(1-z)\left [ 1+a^0_2\; C_2^{3/2}(t)\right ]P_1(2\zeta-1),\nonumber\\
\phi^s(z,\zeta,w^2)&=& \frac{3F_s(s)}{2\sqrt{2N_c}}(1-2z)\left [1+a^s_2\; (1-10z+10z^2) \right ]P_1(2\zeta-1),\nonumber\\
\phi^t(z,\zeta,w^2)&=& \frac{3F_t(s)}{2\sqrt{2N_c}}(1-2z)^2\left [1+a^t_2\; C_2^{3/2}(t)\right ]P_1(2\zeta-1),
\label{eq:phi0st}
\end{eqnarray}
where $p_1$ and $p_2$ denote the momentum of the pion pair, and $p=p_1+p_2$ is the momentum of the
$\rho'$ or $\rho''$ meson. The parameter $z$ is the momentum fraction of the pion pair
and $\zeta$ denotes the momentum fraction for one pion among the pion pair,
while $s=w^2=p^2$ denotes the invariant mass squared of the pion pair.
The Gegenbauer polynomial $C_2^{3/2}(t)=\frac{3}{2}(5t^2-1)$ and $t=2z-1$, and the Legendre polynomial $P_1(2\zeta-1)=2\zeta-1$.

Based on the {\it BABAR} Collaboration analysis of $e^+ e^- \to \pi^+ \pi^-(\gamma)$ data, the form factor $F_\pi$ has been chosen as the form of ~\cite{prd86-032013}
\begin{eqnarray}
F_\pi(s) &=& \frac{1}{1+\sum_i c_i}\cdot \left \{ {\rm GS}_\rho (s, m_\rho, \Gamma_\rho)
\frac{1+c_\omega {\rm BW}_\omega(s, m_\omega, \Gamma_\omega)}{1+c_\omega}
+ \sum_i c_i   {\rm GS}_i   (s, m_i,  \Gamma_i) \right\}, ~~\label{eq:fpiw2}
\end{eqnarray}
with
\beq
{\rm BW}_{\omega}  (s, m, \Gamma)&=& \frac{m^2}{m^2-s-im\Gamma }, \non
{\rm GS}_{\rho,i} (s, m, \Gamma) &=& \frac{m^2\left [ 1+ d(m)\; \Gamma /m \right]}{m^2-s+f(s,m,\Gamma)
-im\Gamma(s,m,\Gamma) }. ~~\label{eq:fpiw3}
\eeq
In the above formulas, ${\rm BW}_{\omega} (s, m, \Gamma)$ is the Breit-Wigner (BW) function~\cite{pr49-519} for the
$\omega$ meson,  while ${\rm GS}_{\rho,i}(s, m, \Gamma)$ are the functions for the $\rho$ meson and its excited states
$i = (\rho', \rho'', \rho''')$ as described by the Gounaris-Sakurai(GS) model based on the
BW model ~\cite{pr49-519,prl21-244}.
The explicit expressions of the functions and relevant parameters in Eqs.~(\ref{eq:fpiw2}) and (\ref{eq:fpiw3})
can be found in Ref.~\cite{prd86-032013}.
In this work, we  single out the component for $\rho'$  and $\rho''$ from the form factors as defined in
Eq.~(\ref{eq:fpiw2}).
We here choose the Gegenbauer moments
\beq
a^0_2=0.30\pm 0.05, \quad a^s_2=0.70\pm 0.20, \quad
a^t_2=-0.40\pm 0.10, \label{eq:gms}
\eeq
by fitting the available experimental data for the decays $B \to P \rho \to P \pi\pi$~\cite{prd95-056008}
where $P$ represents the light pseudoscalar mesons $\pi, K, \eta$, or $\eta^\prime$,
which are slightly different from those determined from the decay $B \to K \rho \to K \pi\pi$ ~\cite{plb763-29}.

For the decays $B_{(s)} \to D (\rho',\rho'') \to D \pi \pi$, the differential decay rate can be written as
\begin{eqnarray}
\frac{d{\cal B}}{dw^2}=\tau_{B}\frac{|\vec{p}_\pi|
|\vec{p}_D | }{32\pi^3m^3_{B}}|{\cal A}|^2,
\label{expr-br}
\end{eqnarray}
where $\tau_{B}$ is the mean lifetime of the $B$ meson, and $|\vec{p}_\pi|$ and $|\vec{p}_D|$ denote the magnitudes of
the $\pi$ and $D$ momenta in the center-of-mass frame of the pion pair,
\begin{eqnarray}
|\vec{p}_\pi|&=&\frac12\sqrt{w^2-4m^2_{\pi}}, \non
|\vec{p}_D|&=&\frac{1}{2} \sqrt{[(m^2_B-m^2_D)^2-2(m^2_B+m^2_D)w^2+w^4]/w^2}.
\end{eqnarray}
The analytic formulas for the corresponding decay amplitudes and relevant functions for the considered decays
$B \to D (\rho',\rho'') \to D \pi \pi$ are the same in form as those given in Ref.~\cite{ma17}
for the cases of $B_{(s)} \to D \rho \to D \pi\pi $ decays.

\section{Numerical results}\label{sec:3}

Besides those Gegenbauer moments in Eq.~(\ref{eq:gms}), the following input parameters ~\cite{PDG2016}
(the masses, decay constants and QCD scale are in units of GeV) are used in the numerical calculations:
\begin{eqnarray}
\Lambda^{(f=4)}_{ \overline{MS} }&=&0.25, \quad m_B=5.280, \quad m_{B_s}=5.367,
\quad m_{D^\pm}=1.870,\quad m_{D^0/\bar{D}^0}=1.865, \nonumber\\
m_{D_s^\pm}&=&1.968,\quad m_{\pi^\pm}=0.140, \quad m_{\pi^0}=0.135, \quad
m_{b}=4.8, \quad m_c=1.27,  \nonumber\\
f_B&=& 0.19, \quad f_{B_s}=0.236, \quad f_D= 0.2119, \quad\quad f_{D_s}=0.249,\quad \nonumber\\
\tau_{B^0}&=&1.520\; {\rm ps}, \quad \tau_{B^+}=1.638\; {\rm ps},
 \quad\tau_{B_{s}}=1.510\; {\rm ps},  \label{eq:inputs}
\end{eqnarray}
and the Wolfenstein parameters
$A=0.811 \pm 0.026,~\lambda=0.22506\pm 0.00050$,~$\bar{\rho} = 0.124_{-0.018}^{+0.019},~\bar{\eta}= 0.356\pm 0.011$.

\begin{table}[htb]
\begin{center}
\caption{The PQCD predictions for the branching ratios of  the quasi-two-body decays
$B_{(s)} \to  D \rho' \to D\pi \pi$ and the two-body decays $B_{(s)} \to  D \rho'$.}
\label{dpda}
\begin{tabular}{ c  c  c } \hline\hline
{\rm Decay modes} & {\rm Quasi-two-body decays} & {\rm Two-body decays}  \\ \hline\hline
$B_{(s)}\to \bar{D}_{(s)}\rho' \to \bar{D}_{(s)}\pi\pi~~~~~~$&${\cal B}$&${\cal B}$ \\ \hline
$B^+\to \bar{D}^0\rho'^+\to \bar{D}^0\pi^+ \pi^0~~~~~~$ &
$(8.68^{+4.84}_{-2.91}(\omega_B)^{+0.42}_{-0.33}(a^t_2)^{+0.11}_{-0.09}(a^0_2)^{+0.04}_{-0.05}(a^s_2)^{+0.65}_{-0.58}(C_{D}))\times10^{-4}~~~~~~$&$(8.65^{+4.88}_{-2.98})\times10^{-3}$ \\
$B^0\to  D^-\rho'^+ \to D^-\pi^+ \pi^0 ~~~~~~$ &
$(6.80^{+4.27}_{-2.49}(\omega_B)^{+0.17}_{-0.12}(a^t_2)^{+0.09}_{-0.03}(a^0_2)^{+0.08}_{-0.08}(a^s_2)^{+0.59}_{-0.50}(C_{D})) \times10^{-4}~~~~~~$&$(6.77^{+4.30}_{-2.53}) \times10^{-3}$ \\
$ B^0\to \bar{D}^0\rho'^0 \to \bar{D}^0\pi^+\pi^-~~~~~~$ &$(9.04^{+3.71}_{-2.75}(\omega_B)^{+4.83}_{-4.26}(a^t_2)^{+0.45}_{-0.59}(a^0_2)^{+0.04}_{-0.07}(a^s_2)^{+0.14}_{-0.10}(C_{D}))\times10^{-6}~~~~~~$&$(9.00^{+6.08}_{-5.18})\times10^{-5}$ \\
$ B_s^0\to D^-\rho'^+ \to D^-\pi^+\pi^0~~~~~~ $ &
$ (4.21^{+0.55}_{-0.61}(\omega_B)^{+1.10}_{-0.81}(a^t_2)^{+0.27}_{-0.25}(a^0_2)^{+0.46}_{-0.39}(a^s_2)^{+0.11}_{-0.19}(C_{D}))\times10^{-7}~~~~~~$ &$(4.19^{+1.34}_{-1.13})\times10^{-6}$  \\
$ B_s^0\to \bar{D}^0\rho'^0 \to \bar{D}^0\pi^+\pi^-~~~~~~$ &
$(1.88^{+0.48}_{-0.20}(\omega_B)^{+0.57}_{-0.34}(a^t_2)^{+0.12}_{-0.11}(a^0_2)^{+0.25}_{-0.17}(a^s_2)^{+0.10}_{-0.08}(C_{D}))\times10^{-7}~~~~~~$&$(1.87^{+0.80}_{-0.45})\times10^{-6}$ \\
$B_s^0\to D_s^- \rho'^+\to D_s^-\pi^+\pi^0~~~~~~$ &
$(5.33^{+2.96}_{-1.80}(\omega_B)^{+0.00}_{-0.00}(a^t_2)^{+0.02}_{-0.01}(a^0_2)^{+0.00}_{-0.00}(a^s_2)^{+0.41}_{-0.40}(C_{D}))\times10^{-4}~~~~~~$ &
$(5.31^{+2.98}_{-1.83})\times10^{-3}$ \\ \hline
$ B_{(s)}\to D_{(s)} \rho' \to D_{(s)}\pi\pi~~~~~~$ & $ {\cal B}$ &${\cal B}$\\ \hline
$ B^+\to D^0\rho'^+\to D^0\pi^+ \pi^0~~~~~~$ &
$(1.51 ^{+0.33}_{-0.29}(\omega_B)^{+0.14}_{-0.05}(a^t_2)^{+0.13}_{-0.07}(a^0_2)^{+0.29}_{-0.25}(a^s_2)^{+0.04}_{-0.04}(C_{D})) \times10^{-8}~~~~~~$ &
$(1.50 ^{+0.48}_{-0.39}) \times10^{-7}$ \\
$ B^+\to D^+\rho'^0\to D^+\pi^+\pi^-~~~~~~ $ &
$( 5.88  ^{+0.90}_{-0.82}(\omega_B)^{+1.46}_{-1.17}(a^t_2)^{+0.07}_{-0.06}(a^0_2)^{+0.88}_{-0.82}(a^s_2)^{+0.05}_{-0.04}(C_{D}))\times10^{-8}~~~~~~$ &
$(5.86^{+1.92}_{-1.65})\times10^{-7}$  \\
$ B^0\to D^0\rho'^0\to D^0\pi^+\pi^-~~~~~~ $ &
$ (9.75 ^{+3.30}_{-3.18}(\omega_B)^{+4.05}_{-2.36}(a^t_2)^{+1.25}_{-1.26}(a^0_2)^{+5.19}_{-3.71}(a^s_2)^{+1.22}_{-0.81}(C_{D}))\times10^{-10}~~~~~~$&
$ (9.71 ^{+7.53}_{-5.61})\times10^{-9} $ \\
$  B^0\to D^+\rho'^-\to D^+\pi^-\pi^0~~~~~~$ &
$ (7.10 ^{+1.06}_{-1.02}(\omega_B)^{+2.61}_{-2.03}(a^t_2)^{+0.03}_{-0.01}(a^0_2)^{+1.32}_{-1.22}(a^s_2)^{+0.13}_{-0.12}(C_{D}))\times10^{-8}~~~~~~$&
$(7.07 ^{+3.10}_{-2.57})\times10^{-7}$  \\
$ B^+\to D_s^+\rho'^0\to D_s^+\pi^+\pi^-~~~~~~ $ &
$(1.38 ^{+0.20}_{-0.20}(\omega_B)^{+0.42}_{-0.34}(a^t_2)^{+0.04}_{-0.04}(a^0_2)^{+0.22}_{-0.20}(a^s_2)^{+0.01}_{-0.01}(C_{D})) \times10^{-6}~~~~~~$&
$(1.37 ^{+0.51}_{-0.44})\times10^{-5}$  \\
$ B^0\to D_s^+\rho'^-\to D_s^+\pi^-\pi^0~~~~~~ $ &
$(2.56 ^{+0.38}_{-0.36}(\omega_B)^{+0.79}_{-0.60}(a^t_2)^{+0.08}_{-0.08}(a^0_2)^{+0.38}_{-0.40}(a^s_2)^{+0.02}_{-0.02}(C_{D}))\times10^{-6}~~~~~~$&
$(2.55 ^{+0.95}_{-0.81})\times10^{-5}$  \\
$ B_s^0\to D^0\rho'^0\to D^0\pi^+\pi^-~~~~~~$ &
$(3.26  ^{+0.47}_{-0.51}(\omega_B)^{+0.29}_{-0.31}(a^t_2)^{+0.21}_{-0.25}(a^0_2)^{+0.07}_{-0.08}(a^s_2)^{+0.19}_{-0.19}(C_{D}))\times10^{-8}~~~~~~$&
$(3.25  ^{+0.62}_{-0.68})\times10^{-7}$  \\
$ B_s^0\to D^+\rho'^-\to D^+\pi^-\pi^0~~~~~~$ &
$(6.56 ^{+0.93}_{-1.03}(\omega_B)^{+0.56}_{-0.66}(a^t_2)^{+0.39}_{-0.53}(a^0_2)^{+0.14}_{-0.18}(a^s_2)^{+0.38}_{-0.38}(C_{D}))\times10^{-8}~~~~~~$&
$(6.53 ^{+1.22}_{-1.40})\times10^{-7}$ \\ \hline\hline
\end{tabular} \end{center}
\end{table}

\begin{table}[htb]
\begin{center}
\caption{The PQCD predictions for the branching ratios of the quasi-two-body decays
$B_{(s)} \to  D \rho'' \to D\pi \pi$  and  the two-body decays $B_{(s)} \to  D \rho''$.}
\label{dpdc}
\begin{tabular}{c c  c } \hline\hline
{\rm Decay modes} & {\rm Quasi-two-body decays} & {\rm Two-body decays}  \\ \hline\hline
$B_{(s)}\to \bar{D}_{(s)}\rho'' \to \bar{D}_{(s)}\pi\pi~~~~~~$&${\cal B}$&${\cal B}$ \\ \hline
 $B^+\to \bar{D}^0\rho''^+\to \bar{D}^0\pi^+ \pi^0~~~~~~$ &
 $(4.58^{+2.62}_{-1.59}(\omega_B)^{+0.17}_{-0.21}(a^t_2)^{+0.06}_{-0.05}(a^0_2)^{+0.01}_{-0.01}(a^s_2)^{+0.29}_{-0.30}(C_{D}))\times 10^{-4}~~~~~~$
 &$(5.65^{+3.26}_{-2.01})\times 10^{-3}$ \\
 $ B^0\to  D^-\rho''^+ \to D^-\pi^+ \pi^0~~~~~~$ &
 $(3.30^{+2.09}_{-1.21}(\omega_B)^{+0.08}_{-0.07}(a^t_2)^{+0.03}_{-0.02}(a^0_2)^{+0.05}_{-0.04}(a^s_2)^{+0.26}_{-0.26}(C_{D})) \times 10^{-4}~~~~~~$&
 $(4.07^{+2.60}_{-1.53}) \times 10^{-3}$ \\
 $ B^0\to \bar{D}^0\rho''^0 \to \bar{D}^0\pi^+\pi^-~~~~~~$ &
 $(5.68^{+2.14}_{-1.65}(\omega_B)^{+2.96}_{-2.46}(a^t_2)^{+0.09}_{-0.09}(a^0_2)^{+0.27}_{-0.34}(a^s_2)^{+0.09}_{-0.07}(C_{D}))\times 10^{-6}~~~~~~$&
 $(7.00^{+4.51}_{-3.98})\times 10^{-5}$ \\
 $ B_s^0\to D^-\rho''^+ \to D^-\pi^+\pi^0~~~~~~$ &
 $ (2.08^{+0.49}_{-0.43}(\omega_B)^{+0.78}_{-0.60}(a^t_2)^{+0.11}_{-0.13}(a^0_2)^{+0.34}_{-0.30}(a^s_2)^{+0.04}_{-0.03}(C_{D}))\times 10^{-7}~~~~~~$ &
 $(2.56^{+1.21}_{-0.97})\times 10^{-6}$  \\
 $ B_s^0\to \bar{D}^0\rho''^0 \to \bar{D}^0\pi^+\pi^-~~~~~~$ &
 $(1.04^{+0.23}_{-0.21}(\omega_B)^{+0.39}_{-0.31}(a^t_2)^{+0.06}_{-0.07}(a^0_2)^{+0.17}_{-0.16}(a^s_2)^{+0.02}_{-0.02}(C_{D}))\times 10^{-7}~~~~~~$ &
$(1.28^{+0.60}_{-0.51})\times 10^{-6}$\\
 $ B_s^0\to D_s^- \rho''^+\to D_s^-\pi^+\pi^0~~~~~~$ &
  $(2.57^{+1.46}_{-0.89}(\omega_B)^{+0.00}_{-0.00}(a^t_2)^{+0.01}_{-0.01}(a^0_2)^{+0.00}_{-0.00}(a^s_2)^{+0.20}_{-0.19}(C_{D}))\times 10^{-4}~~~~~~$&
$(3.17^{+1.82}_{-1.11})\times 10^{-5}$  \\ \hline
  $B_{(s)}\to D_{(s)} \rho'' \to D_{(s)}\pi\pi~~~~~~$&${\cal B}$&${\cal B}$\\ \hline
  $ B^+\to D^0\rho''^+\to D^0\pi^+ \pi^0~~~~~~$ &
$(8.39 ^{+1.17}_{-1.38}(\omega_B)^{+1.41}_{-0.89}(a^t_2)^{+0.64}_{-0.55}(a^0_2)^{+1.68}_{-1.27}(a^s_2)^{+0.06}_{-0.22}(C_{D}) )\times 10^{-9}~~~~~~$ &
$(1.03 ^{+0.31}_{-0.27} )\times 10^{-7}$ \\
 $ B^+\to D^+\rho''^0\to D^+\pi^+\pi^-~~~~~~$ &
 $ (1.55  ^{+0.07}_{-0.07}(\omega_B)^{+0.36}_{-0.17}(a^t_2)^{+0.01}_{-0.01}(a^0_2)^{+0.33}_{-0.29}(a^s_2)^{+0.02}_{-0.02}(C_{D}))\times 10^{-8}~~~~~~$ &
$(1.91  ^{+0.61}_{-0.43})\times 10^{-7}$  \\
 $ B^0\to D^0\rho''^0\to D^0\pi^+\pi^-~~~~~~$ &
  $(  3.62 ^{+0.90}_{-1.18}(\omega_B)^{+1.58}_{-0.81}(a^t_2)^{+0.45}_{-0.59}(a^0_2)^{+2.46}_{-1.79}(a^s_2)^{+0.25}_{-0.42}(C_{D}))\times 10^{-10}~~~~~~$  &
$ (4.46 ^{+3.82}_{-2.97})\times 10^{-9}$ \\
 $  B^0\to D^+\rho''^-\to D^+\pi^-\pi^0~~~~~~$ &
 $ ( 1.41 ^{+0.06}_{-0.04}(\omega_B)^{+0.73}_{-0.37}(a^t_2)^{+0.01}_{-0.03}(a^0_2)^{+0.36}_{-0.29}(a^s_2)^{+0.03}_{-0.04}(C_{D}))\times 10^{-8}~~~~~~$&
$(1.74 ^{+1.01}_{-0.59})\times 10^{-7}$  \\
 $ B^+\to D_s^+\rho''^0\to D_s^+\pi^+\pi^-~~~~~~$ &
 $(3.25 ^{+0.02}_{-0.14}(\omega_B)^{+1.32}_{-0.77}(a^t_2)^{+0.08}_{-0.08}(a^0_2)^{+0.61}_{-0.55}(a^s_2)^{+0.03}_{-0.03}(C_{D})) \times 10^{-7}~~~~~~$ &
$(4.01 ^{+1.80}_{-1.18} )\times 10^{-6}$ \\
 $ B^0\to D_s^+\rho''^-\to D_s^+\pi^-\pi^0~~~~~~$ &
 $(6.03 ^{+0.02}_{-0.26}(\omega_B)^{+2.44}_{-1.44}(a^t_2)^{+0.15}_{-0.14}(a^0_2)^{+1.14}_{-1.02}(a^s_2)^{+0.06}_{-0.05}(C_{D}))\times 10^{-7}~~~~~~$  &
$(7.44 ^{+3.33}_{-2.21})\times 10^{-6}$\\
 $ B_s^0\to D^0\rho''^0\to D^0\pi^+\pi^-~~~~~~$ &
 $(1.65  ^{+0.32}_{-0.26}(\omega_B)^{+0.20}_{-0.15}(a^t_2)^{+0.14}_{-0.10}(a^0_2)^{+0.06}_{-0.05}(a^s_2)^{+0.09}_{-0.08}(C_{D}))\times 10^{-8}~~~~~~$ &
$(2.04  ^{+0.52}_{-0.41})\times 10^{-7}$ \\
 $ B_s^0\to D^+\rho''^-\to D^+\pi^-\pi^0~~~~~~$ &
 $(3.31 ^{+0.64}_{-0.52}(\omega_B)^{+0.40}_{-0.30}(a^t_2)^{+0.26}_{-0.20}(a^0_2)^{+0.13}_{-0.08}(a^s_2)^{+0.18}_{-0.17}(C_{D}))\times 10^{-8}~~~~~~$ &
$(4.08 ^{+0.99}_{-0.80})\times 10^{-7}$ \\ \hline\hline
\end{tabular} \end{center}
\end{table}

In the second columns of Tables~\ref{dpda} and~\ref{dpdc}, we present  the PQCD predictions for the branching ratios of
the quasi-two-body decays $B_{(s)} \to  D (\rho',\rho'') \to D \pi \pi$.
The main errors come from  the uncertainties of the input
parameters in the wave functions of the $B_{(s)}$ meson and the final state mesons:
$\omega_B = 0.40 \pm 0.04$ and $\omega_{B_s}=0.50 \pm 0.05$, $a^t_2=
-0.40\pm0.10$, $a^0_2=0.30\pm 0.05$ and $a^s_2=0.70\pm0.20$, $C_{D}=0.5\pm 0.1$ and $C_{D_s}=0.4\pm 0.1$, respectively.

As a special feature of our PQCD framework, we can  extract the branching ratios for the two-body decays
$B_{(s)} \to  D (\rho',\rho'')$ from the corresponding quasi-two-body decays if one knows the decay
rates of $(\rho',\rho'')\to\pi\pi$ transitions reliably.
In Ref.~\cite{plb763-29}, the authors found
\beq
{\cal B}(\rho'\to\pi\pi)=\Gamma_{\rho'\to\pi\pi}/\Gamma_{\rho'}
=\left (10.04^{+5.23}_{-2.61} \right ) \times 10^{-2},
\label{eq:br31}
\eeq
by using the formula
\begin{eqnarray}
\Gamma_{\rho'\to\pi\pi}=\frac{g^2_{\rho'\pi\pi}}{6\pi}\frac{|\overrightarrow{p_\pi}(m^2_{\rho'})|^3}{m^2_{\rho'}}
\label{bf}
\end{eqnarray}
and the measured value of $\Gamma_{\rho'}=0.311\pm 0.062$ GeV~\cite{zpc62-455}.
The value of ${\cal B}(\rho'\to\pi\pi)\approx 10\%$ \cite{plb763-29}
is also consistent with the range $[4.56\%,10.0\%]$ as predicted in Refs.~\cite{ijmpa13-5443,zpc62-455}.
By using the same method, we find $f_{\rho''}=0.103^{+0.011}_{-0.012}$ GeV \cite{1704.07566}
when $\Gamma_{\rho''\to e^+e^-}=0.69\pm 0.15$ keV~\cite{zpc62-455} is adopted.
Again we find \cite{1704.07566}
\beq
{\cal B}(\rho''\to\pi\pi)=\left (8.11^{+2.22}_{-1.47}\right )\times 10^{-2}. \label{eq:br32}
\eeq
Of course, we know that the resonance parameters for $\rho''$ are still not well determined ~\cite{prd88-093002};
more theoretical studies  and experimental measurements are indeed required to improve
the estimation for those parameters.

By using the simple relation between the decay rate of the quasi-two-body decay and the
corresponding two-body ones
\beq
{\cal B}( B_{(s)} \to  D (\rho', \rho'') \to D \pi \pi ) =
{\cal B}( B_{(s)} \to  D (\rho', \rho'') ) \cdot {\cal B}((\rho', \rho'') \to \pi \pi),
\label{eq:def1}
\eeq
one can extract the branching ratios ${\cal B}(B_{(s)} \to  D \rho') $ and ${\cal B}(B_{(s)} \to  D \rho'')$
from the PQCD predictions for the branching ratios
of those quasi-two-body decays $B_{(s)} \to  D (\rho', \rho'') \to D \pi \pi$, if we take ${\cal B}(\rho'\to\pi\pi)$
and ${\cal B}(\rho''\to\pi\pi)$ as given in Eqs.~(\ref{eq:br31}) and (\ref{eq:br32}) as input.
In the last column of Tables \ref{dpda} and \ref{dpdc}, we listed the PQCD predictions
for ${\cal B}(B_{(s)} \to  D \rho') $ and
${\cal B}(B_{(s)} \to  D \rho'')$, where the individual errors have been added in quadrature.

\begin{figure}[tbp]
\centerline{\epsfxsize=8.5cm \epsffile{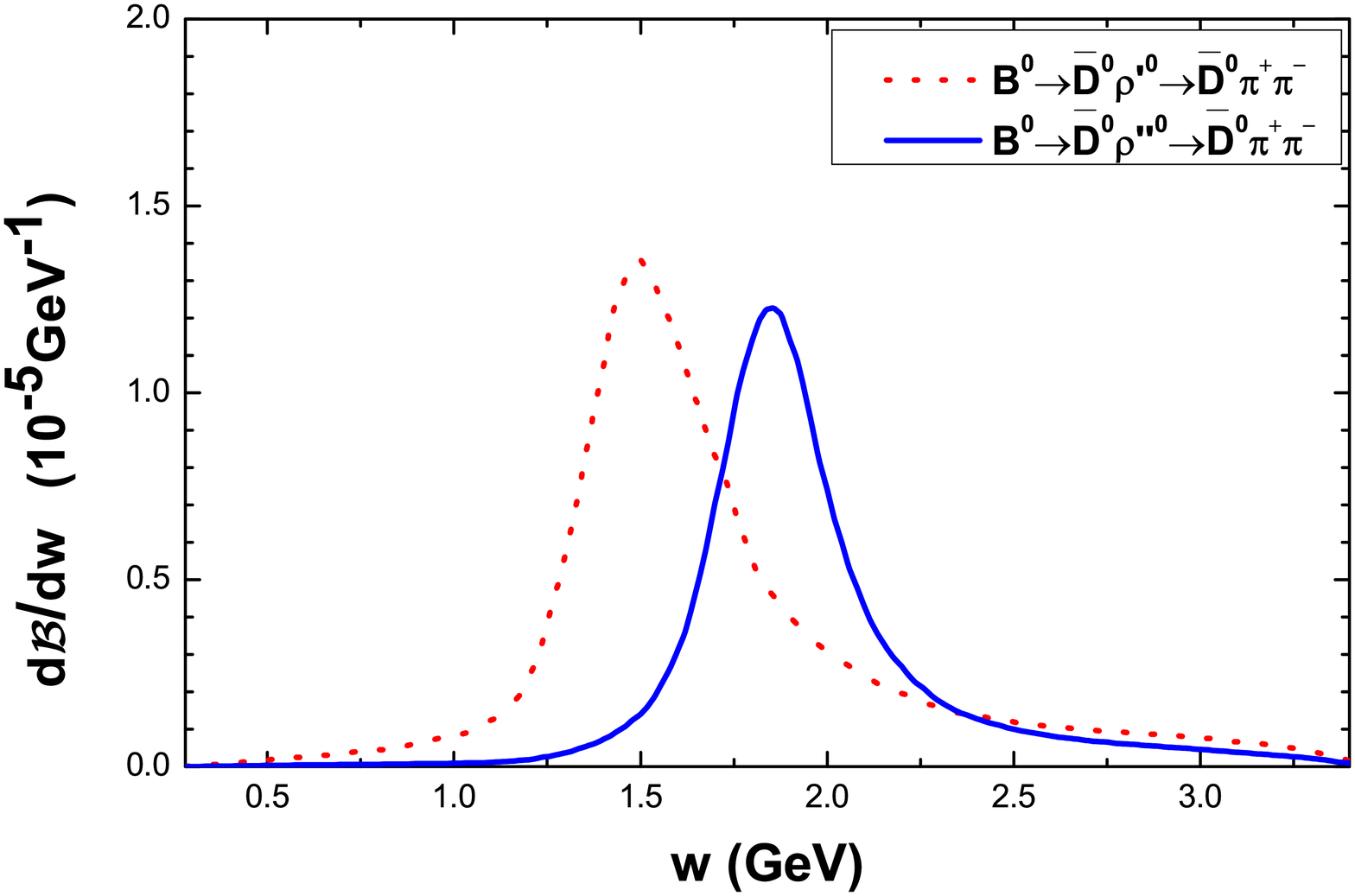}
            \epsfxsize=8.5cm \epsffile{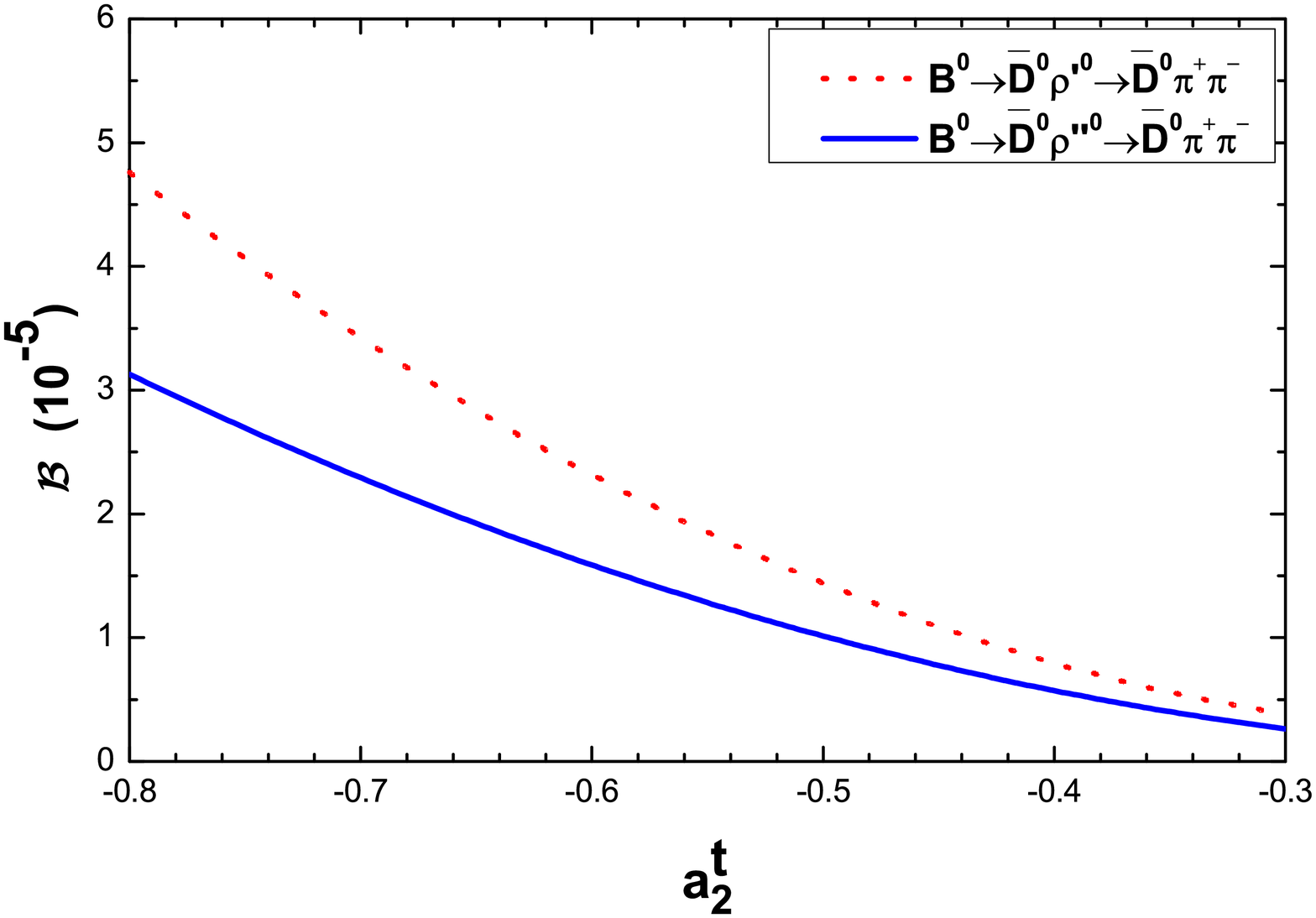}}
\vspace{-0.2cm}
  {\scriptsize\bf (a)\hspace{7.9cm}(b)}
\caption{(a) The differential branching ratios for the ${\cal B}(B^0\to \bar{D}^0 (\rho'^0,\rho''^0) \to \bar{D}^0\pi^+\pi^-)$ decays.
(b) The branching ratio of
${\cal B}(B^0\to \bar{D}^0 (\rho'^0,\rho''^0) \to \bar{D}^0\pi^+\pi^-)$ decays with $a^t_2=[-0.8,-0.3]$. }
\label{fig:fig1}
\end{figure}

From our studies and the PQCD predictions as listed in above tables, we have the following observations:
\begin{itemize}
\item[(1)]
Unlike the fixed kinematics of the two-body $B_{(s)}$ meson decays, the decay amplitudes
of the quasi-two-body $B_{(s)}$ meson decays considered in this paper do have a strong dependence on the $\pi^+\pi^-$
invariant mass $s=w^2$.
In Fig.~\ref{fig:fig1}(a), we plot the $w$-dependence of the differential decay rates for
$B^0\to \bar{D}^0\rho'^0 \to \bar{D}^0\pi^+\pi^-$ (the red dots curve)
and $B^0\to \bar{D}^0\rho''^0 \to \bar{D}^0\pi^+\pi^-$ (the blue solid curve).
As discussed in Refs.~\cite{ma17,plb763-29},  the main contribution to the branching ratios lies
in the region around the pole mass of the resonance $m_{\rho'}=1.45$ GeV and  $m_{\rho''}=1.7$ GeV.
Numerically,  ${\cal B}(B^0\to \bar{D}^0\rho'^0 \to \bar{D}^0\pi^+\pi^-)$ is a little larger
than ${\cal B}(B^0\to \bar{D}^0\rho''^0 \to \bar{D}^0\pi^+\pi^-)$, since the relevant parameters such as $c_{\rho^\prime}$
and $c_{\rho^{\prime\prime}}$ are a little different for the decay involving
$\rho^\prime$ or $\rho^{\prime\prime}$.
Such kinds of differences also exist for other decay channels; one can easily find them from the values as listed
in Tables~\ref{dpda} and~\ref{dpdc}.

\item[(2)]
Our prediction for the central value of ${\cal B}(B^0\to \bar{D}^0\rho'^0 \to \bar{D}^0\pi^+\pi^-)
\approx 0.90 \times 10^{-5}$ is less than the experimental result reported by LHCb~\cite{prd92-032002}:
$1.36~(1.91) \times 10^{-5}$ in the Isobar model (K-matrix model).
For $B^0\to \bar{D}^0\rho''^0 \to \bar{D}^0\pi^+\pi^-$, furthermore,  our prediction is
${\cal B}(B^0\to \bar{D}^0\rho''^0 \to \bar{D}^0\pi^+\pi^-) \approx 0.57 \times 10^{-5}$,
while the measured value from LHCb  was  $0.33~(0.73) \times 10^{-5}$ in the Isobar model (K-matrix model)
~\cite{prd92-032002}.
If we take the still large theoretical errors into account, our PQCD predictions
as listed in Tables~\ref{dpda} and~\ref{dpdc} do agree well with those currently available data.

\item[(3)]
The dominant theoretical error comes from the uncertainty of
$\omega_B = 0.40 \pm 0.04$ and $\omega_{B_s}=0.50 \pm 0.05$: about $20\%-50\%$ of the central values.
The PQCD predictions also have a strong dependence on the magnitude of
the Gegenbauer coefficients, specifically on the value of $a^t_2$.
In Fig.~\ref{fig:fig1}(b), we plot the PQCD predictions for the branching ratios of the decay
${\cal B}(B^0\to \bar{D}^0\rho'^0 \to \bar{D}^0\pi^+\pi^-)$ (the red curve) and
${\cal B}(B^0\to \bar{D}^0 \rho''^0 \to \bar{D}^0\pi^+\pi^-)$ (the blue curve)
in the range of $a^t_2=[-0.8,-0.3]$ (although in this paper, we assume $a^t_2=-0.4\pm0.1$)
while other  parameters take their central values.
For some decay modes, we observe similar strong $a^t_2$-dependence, as listed in Table~\ref{dpda},
\begin{eqnarray}
{\cal B}(B_s^0\to D^-\rho'^+ \to D^-\pi^+\pi^0) &=&4.21^{+1.10}_{-0.81}(a^t_2) \times 10^{-7}, \non
{\cal B}(B_s^0\to \bar{D}^0\rho'^0 \to \bar{D}^0\pi^+\pi^-) &=&1.88^{+0.57}_{-0.34}(a^t_2)\times 10^{-7}, \non
{\cal B}(B^+\to D^+\rho'^0\to D^+\pi^+\pi^-) &=&5.88^{+1.46}_{-1.17}(a^t_2)\times 10^{-8}, \non
{\cal B}(B^0\to D^0\rho'^0\to D^0\pi^+\pi^-) &=& 9.75^{+4.05}_{-2.36}(a^t_2) \times 10^{-10}, \non
{\cal B}( B^+\to D_s^+\rho'^0\to D_s^+\pi^+\pi^-)&=&1.38^{+0.42}_{-0.34}(a^t_2)\times 10^{-6}, \non
{\cal B}(B^0\to D_s^+\rho'^-\to D_s^+\pi^-\pi^0) &=&2.56^{+0.79}_{-0.60}(a^t_2)\times 10^{-6}.
\label{eq:bd20}
\end{eqnarray}
It is easy to see that the theoretical error due to $a_2^t=-0.4\pm 0.1$ amounts to $20\%-40\%$ to the
central values for the decays in Eq.~(\ref{eq:bd20}). For other remaining decays, the corresponding
error due to $a_2^t$ is only about $ 10 \%$.
The same situation appears for the considered $B_{(s)} \to D \rho'' \to D \pi \pi$ decays.

\item[(4)]
We find a new way to estimate the decay rates of the two-body decays $B_{(s)} \to D (\rho', \rho'')$.
The PQCD predictions for  ${\cal B}(B_{(s)} \to  D (\rho', \rho''))$  are listed in the third column of
Tables~\ref{dpda} and~\ref{dpdc}.
When compared with the numerical results for ${\cal B}(B_{(s)} \to  D \rho )$ decays as listed in
Tables I and II of Ref.~\cite{ma17}, we find that
the PQCD predictions for the branching ratios of the similar decay modes but involving different
$\rho$ or $\rho'$ as one of the two final state mesons are similar in magnitudes:  for example,
${\cal B}(B^+\to \bar{D}^0\rho'^+)\approx 0.87 \times 10^{-2}$ vs ${\cal B}(B^+\to \bar{D}^0\rho^+)=1.15 \times 10^{-2}$,
and ${\cal B}(B^+\to  D^+\rho'^0)=5.86 \times 10^{-7}$  vs ${\cal B}(B^+\to  D^+\rho^0)=5.33 \times 10^{-7}$.

\end{itemize}

\section{Summary}\label{sec:4}

In this paper, we calculated the branching ratios of the quasi-two-body $B_{(s)} \to  D (\rho', \rho'')
\to D \pi \pi$ decays by employing the PQCD factorization approach.
The contributions from the $\rho'$ and $\rho''$ resonant states were singled out from the given
timelike form factor $F_\pi$ in the $P$-wave two-pion distribution amplitudes $\Phi^{I=1}_{\pi\pi}$.
With the estimated branching fraction for $\rho' \to \pi\pi$ and $\rho'' \to \pi\pi$ ,
we have also extracted the theoretical predictions for the
branching ratios for the two-body decays $B_{(s)} \to  D \rho' $ and $B_{(s)} \to  D \rho'' $.
From the analytical and numerical calculations, we found the following points:
\begin{itemize}
\item[(1)]
The PQCD predictions for the branching ratios of the considered quasi-two-body decays
$B_{(s)} \to D (\rho',\rho'') \to D \pi \pi $ are in the range of $10^{-10} \sim 10^{-4}$. Those decay channels with
large decay rate, say $\geq 10^{-6}$,  could be measured and tested at the future LHCb and Belle-II experiments.

\item[(2)]
The PQCD predictions for ${\cal B}(B^0\to \bar{D}^0 (\rho'^0,\rho''^0) \to \bar{D}^0 \pi^+\pi^-)$
agree well with the measured values as reported by LHCb if one takes still large theoretical
errors into account.

\item[(3)]
One can extract the decay rates for the two-body decays $B_{(s)} \to  D (\rho',\rho'')$  from the
PQCD predictions for the branching ratios of the corresponding quasi-two-body decays
$B_{(s)} \to  D (\rho',\rho'')\to D \pi\pi$.

\item[(4)]
The PQCD predictions for ${\cal B}(B_{(s)} \to  D \rho')$ and ${\cal B}(B_{(s)} \to  D \rho)$
are similar in magnitude. It is an interesting relation to be tested by the future experimental measurements.

\end{itemize}

\begin{acknowledgments}

Many thanks to Hsiang-nan Li for valuable discussions.
This work is supported by the National Natural Science Foundation of China under Grants No.~11235005 and No.~11547038.
Ai-Jun Ma and Ya Li are also supported by Postgraduate Research \& Practice Innovation Program of Jiangsu Province under Grants No.~KYCX17-1056 and No.~KYCX17-1057.

\end{acknowledgments}



\begin{thebibliography}{99}


\bibitem{1612.07233}
Y.~Amhis {\it et al.}  (Heavy Flavor Averaging Group Collaboration), Averages of $b$-hadron, $c$-hadron, and  $\tau$-lepton
properties as of summer 2016, arXiv:1612.07233v2.

\bibitem{plb553-159}
A.~Satpathy {\it et al.}  (Belle Collaboration), Study of $\bar{B}^0 \to D^{(*)0} \pi^+ \pi^-$ decays,
\plb  {\bf 553}, 159 (2003).

\bibitem{prd76-012006}
A.~Kuzmin {\it et al.}  (Belle Collaboration), Study of $\bar{B}^0 \to D^0 \pi^+ \pi^-$ decays,
\prd {\bf 76}, 012006 (2007).

\bibitem{prd79-112004}
B.~Aubert {\it et al.}  ({\it BABAR} Collaboration), Dalitz-plot analysis of $B^- \to D^+ \pi^- \pi^-$,
\prd {\bf 79}, 112004 (2009).


\bibitem{prd92-032002}
R.~Aaij {\it et al.}  (LHCb Collaboration), Dalitz-plot analysis of $B^0 \to \bar{D}^0 \pi^+ \pi^-$ decays,
\prd {\bf 92}, 032002 (2015).

\bibitem{prd94-072001}
R.~Aaij {\it et al.}  (LHCb Collaboration), Amplitude analysis of $B^- \to D^+ \pi^- \pi^-$ decays,
\prd {\bf 94}, 072001 (2016).

\bibitem{1702.01535}
T.~Gershon (LHCb Collaboration), Current challenges and future prospects for $\gamma$ from $B \to Dhh'$ decays,
Proc. Sci., CKM2016 (2017) 115[arXiv:1702.01535].


\bibitem{pm44-1068}
R.~H.~Dalitz, On the analysis of $\tau$-meson data and the nature of the $\tau$-meson,
Phil.Mag. {\bf 44}, 1068 (1953).

\bibitem{pr135-B511}
G.~N.~Fleming, Recoupling effects in the Isobar model. I. General formalism for three-pion scattering,
\pr  {\bf 135}, B551 (1964).

\bibitem{pr166-1731}
D.~Morgan, Phenomenological analysis of $I=\frac{1}{2}$ single-pion production processes in the energy range 500 to 700 MeV,
\pr  {\bf 166}, 1731 (1968).

\bibitem{prd11-3165}
D.~Herndon, P.~S\"oding, and R.~J.~Cashmore, Generalized Isobar model formalism,
\prd  {\bf 11}, 3165 (1975).

\bibitem{epja16-229}
V.~V.~Anisovich and A.~V.~Sarantsev, K-matrix analysis of the $(IJ^{PC} = 00^{++})$-wave in the mass region below 1900 MeV,
\epja  {\bf 16}, 229 (2003).

\bibitem{prd83-112010}
J.~P.~Lees {\it et al.}  ({\it BABAR} Collaboration), Amplitude analysis of  $B^0 \to K^+\pi^-\pi^0$  and evidence of direct $CP$ violation in  $B \to K^*\pi$ decays,
\prd {\bf 83}, 112010 (2011).


\bibitem{prd79-072006}
B.~Aubert {\it et al.}  ({\it BABAR} Collaboration), Dalitz-plot analysis of $B^{\pm} \to \pi^{\pm}\pi^{\pm}\pi^{\mp}$ decays,
\prd  {\bf 79}, 072006 (2009).

\bibitem{prd78-072006}
M.~Fujikawa {\it et al.} (Belle Collaboration), High-statistics study of the $\tau^- \to \pi^- \pi^0 \nu_\tau$ decay,
\prd {\bf 78}, 072006 (2008).

\bibitem{prd86-032013}
J.~P.~Lees {\it et al.} ({\it BABAR} Collaboration), Precise measurement of the $e^+e^- \to \pi^+\pi^-(\gamma)$ cross section with the initial-state radiation method at {\it BABAR},
\prd  {\bf 86}, 032013 (2012).

\bibitem{PDG2016}
C.~Patrignani  {\it et al.}  (Particle Data Group), Review of particle physics, \cpc  {\bf 40}, 100001 (2016).

\bibitem{plb622-207}
A.~Furman, R.~Kami\'nski, L.~Le\'sniak, and B.~Loiseau, Long-distance effects and final state interactions in
$B \to \pi\pi K$ and $B \to K\bar{K}K$ decays, \plb  {\bf622}, 207 (2005).

\bibitem{prd74-114009}
B.~El-Bennich {\it et al.}, Interference between $f_0(980)$ and $\rho(770)^0$ resonances in $B \to \pi^+ \pi^- K$ decays,
\prd  {\bf74}, 114009 (2006).

\bibitem{prd79-094005}
B.~El-Bennich {\it et al.}, $CP$ violation and kaon-pion interactions in $B \to K \pi^+ \pi^-$ decays,
\prd  {\bf79}, 094005 (2009); , {\bf83}, 039903 (E)(2011).

\bibitem{prd81-094033}
O.~Leitner, J.-P.~Dedonder, B.~Loiseau, and R.~Kami\'nski, $K^*$ resonance effects on direct $CP$ violation in $B \to \pi \pi K$,
\prd  {\bf81}, 094033(2010);, {\bf82}, 119906 (2010)(E).


\bibitem{prd66-054015}
H.~Y.~Cheng and K.~C.~Yang, Nonresonant three-body decays of $D$ and $B$ mesons, \prd  {\bf66}, 054015 (2002).

\bibitem{prd76-094006}
H.~Y.~Cheng, C.~K.~Chua, and A.~Soni, Charmless three-body decays of $B$ mesons, \prd  {\bf76}, 094006 (2007).

\bibitem{prd88-114014}
H.~Y.~Cheng and C.~K.~Chua,
Branching fractions and direct $CP$ violation in charmless three-body decays of $B$ mesons, \prd  {\bf88}, 114014 (2013).

\bibitem{prd89-074025}
H.~Y.~Cheng and C.~K.~Chua, Charmless three-body decays of $B_s$ mesons, \prd  {\bf89}, 074025 (2014).

\bibitem{prd94-094015}
H.~Y.~Cheng, C.~K.~Chua, and Z.~Q.~Zhang, Direct $CP$ violation in charmless three-body decays of $B$ mesons, \prd  {\bf 94}, 094015 (2016).

\bibitem{prd89-094007}
Y.~Li, Comprehensive study of $\bar{B}^0 \to K^0(\bar{K}^0)K^{\mp}\pi^{\pm}$ decays in the factorization approach,
\prd {\bf 89}, 094007 (2014); Y.~Li, Branching fractions and direct $CP$ asymmetries of $\bar{B}^0_s \to K^0 h^+ h^{\prime -}(h^{(\prime)}=K,\pi)$ decays,
Sci. China Phys. Mech. Astron. {\bf58}, 031001 (2015).

\bibitem{npb899-247}
S.~Kr\"ankl, T.~Mannel, and J.~Virto, Three-body nonleptonic $B$ decays and QCD factorization,
\npb{\bf899}, 247 (2015).

\bibitem{epjc75-536}
C.~Wang, Z.~H. Zhang, Z.~Y.~Wang, and X.~H.~Guo, Localized direct $CP$ violation in $B^{\pm}
\to \rho^0(\omega)\pi^{\pm}\to \pi^+\pi^-\pi^{\pm}$, \epjc {\bf75}, 536 (2015).

\bibitem{prd87-076007}
Z.~H.~Zhang, X.~H.~Guo, and Y.~D.~Yang, $CP$ violation in $B^{\pm} \to \pi^{\pm}\pi^+\pi^-$ in the region
with low invariant mass of one $\pi^+\pi^-$ pair, \prd  {\bf87}, 076007 (2013).


\bibitem{prd70-034033}
S.~Fajfer, T.~N.~Pham, and A.~Prapotnik, $CP$ violation in the partial width asymmetries for
$B^- \to \pi^+\pi^-K^-$ and  $B^- \to K^+K^-K^-$ decays, \prd  {\bf70}, 034033 (2004).


\bibitem{plb561-258}
C.~H.~Chen and H.~n.~Li, Three-body nonleptonic $B$ decays in perturbative QCD,
\plb {\bf 561}, 258 (2003).

\bibitem{prd70-054006}
C.~H.~Chen and H.~n.~Li, Vector-pseudoscalar two-meson distribution amplitudes in three body $B$ meson decays,
\prd  {\bf 70}, 054006 (2004).

\bibitem{prd89-074031}
W.~F.~Wang, H.~C.~Hu, H.~n.~Li, and C.~D.~L\"u, Direct $CP$ asymmetries of three-body $B$ decays in perturbative QCD,
\prd {\bf 89}, 074031 (2014).

\bibitem{prd91-094024}
W.~F.~Wang, H.~n.~Li, W.~Wang, and C.~D.~L\"u, S-wave resonance contributions to the $B^0_{(s)} \to J/\psi
\pi^+ \pi^-$ and $B_s \to \pi^+ \pi^-\mu^+\mu^-$ decays, \prd {\bf 91}, 094024 (2015).

\bibitem{epjc76-675}
Y.~Li, A.~J.~Ma, W.~F.~Wang, and Z.~J.~Xiao, The S-wave resonance contributions to the three-body decays $B^0_{(s)} \to
\eta_c f_0(X) \to \eta_c \pi^+ \pi^-$ in perturbative QCD approach, \epjc  {\bf76}, 675 (2016).

\bibitem{plb763-29}
W.~F.~Wang and H.~n.~Li, Quasi-two-body decays $B \to K\rho \to K\pi\pi$ in perturbative QCD approach,
\plb  {\bf763}, 29 (2016).

\bibitem{ma17}
A.~J.~Ma, Y.~Li, W.~F.~Wang, and Z.~J.~Xiao, The quasi-two-body decays $B_{(s)} \to  (D_{(s)},\bar{D}_{(s)})
\rho \to (D_{(s)}, \bar{D}_{(s)})\pi \pi$ in the perturbative QCD factorization approach,
\npb{\bf 923}, 54 (2017).

\bibitem{prd95-056008}
Y.~Li, A.~J.~Ma, W.~F.~Wang, and Z.~J.~Xiao,
Quasi-two-body decays $B_{(s)} \to P\rho \to P\pi\pi$ in perturbative QCD approach, \prd {\bf 95}, 056008 (2017).

\bibitem{1701.01844}
A.~J.~Ma, Y.~Li, W.~F.~Wang, and Z.~J.~Xiao, $S$-wave resonance contributions to the  $B^0_{(s)} \to
\eta_c(2s)\pi^+\pi^-$ in the perturbative QCD factorization approach, \cpc {\bf 41}, 083105 (2017).

\bibitem{1701.02941}
R.~Zhou, Y.~Li, and W.~F.~Wang, The S-wave resonance contributions in the $B^0_s$ decays into $\psi(2S,3S)$
plus pion pair, \epjc {\bf 77}, 199 (2017). 

\bibitem{1704.07566} 	
Y.~Li, A.~J.~Ma, W.~F.~Wang, and Z.~J.~Xiao, Quasi-two-body decays $B_{(s)} \to P\rho^\prime(1450),
P\rho^{\prime\prime}(1700) \to P\pi\pi$ in the perturbative QCD approach, \prd {\bf 96}, 036014 (2017).


\bibitem{prd72-075013}
G.~Engelhard, Y.~Nir, and G.~Raz, SU(3) relations and the $CP$ asymmetry in $B \to K_S K_S K_S$,
\prd  {\bf 72}, 075013 (2005).

\bibitem{prd72-094031}
M.~Gronau and J.~L.~Rosner, Symmetry relations in charmless $B \to PPP$ decays, \prd {\bf 72}, 094031 (2005).

\bibitem{prd84-056002}
M.~Imbeault and D.~London, SU(3) breaking in charmless $B$ decays, \prd  {\bf84}, 056002 (2011).

\bibitem{plb726-337}
B.~Bhattacharya, M.~Gronau, and J.~L.~Rosner, $CP$ asymmetries in three-body $B^\pm$ decays to charged
pions and kaons, \plb  {\bf726}, 337 (2013).

\bibitem{prd89-074043}
B.~Bhattacharya {\it et al.}, Charmless $B \to PPP$ decays: The fully symmetric final state, \prd  {\bf89}, 074043 (2014).

\bibitem{plb728-579}
D.~Xu, G.~N.~Li, and X.~G.~He, $U$-spin analysis of $CP$ violation in $B^-$ decays into three charged light pseudoscalar
mesons, \plb  {\bf728}, 579 (2014).

\bibitem{prd91-014029}
X.~G.~He, G.~N.~Li, and D.~Xu, SU(3) and isospin breaking effects on $B \to PPP$ amplitudes,
\prd  {\bf 91}, 014029 (2015).


\bibitem{prd92-054010}
J.~H.~A.~Nogueira {\it et al.}, $CP$ violation: Dalitz interference, CPT, and final state interactions,
\prd {\bf92}, 054010 (2015).

\bibitem{prd84-034040}
N.~R.~-L.~Lorier, M.~Imbeault, and D.~London,
Diagrammatic analysis of charmless three-body $B$ decays, \prd {\bf 84}, 034040 (2011).

\bibitem{prd60-074029}
J.~L.~Rosner, On large final-state phases in heavy meson decays, \prd  {\bf60}, 074029 (1999).

\bibitem{prd65-096007}
C.~K.~Chua, W.~S.~Hou, and K.~C.~Yang, Final state rescattering and color suppressed $\bar{B}^0 \to D^{(*)0} h^0$ decays,
\prd  {\bf65}, 096007 (2002).

\bibitem{prd69-094018}
Y.~Y.~Keum {\it et al.}, Nonfactorizable contributions to $B \to D^{(*)}$ M decays,
\prd  {\bf69}, 094018 (2004).

\bibitem{prd75-074021}
C.~W.~Chiang and E.~Senaha, Update analysis of two-body charmed $B$ meson decays,
\prd  {\bf75}, 074021 (2007).

\bibitem{prd78-014018}
R.~H.~Li, C.~D.~L\"u, and H.~Zou, $B(B_s) \to D_{(s)}P, D_{(s)}V, D^*_{(s)}P,$ and $D^*_{(s)}V$ decays in the perturbative
QCD approach, \prd  {\bf78}, 014018 (2008).

\bibitem{Grozin}
A.~G.~Grozin, On wave functions of mesonic pairs and mesonic resonances,
Sov.~J.~Nucl.~Phys. {\bf 38},  289 (1983); A.~G.~Grozin, One and two particle wave
functions of multihadron systems, Theor.~Math.~Phys.   {\bf 69}, 1109 (1986).

\bibitem{Muller}
D.~M\"uller {\it et al.}, Wave functions, evolution equations, and evolution kernels
from light-ray operators of QCD, Fortschr. Phys.  {\bf 42}, 101 (1994).

\bibitem{prl81-1782}
M.~Diehl, T.~Gousset, B.~Pire, and O.~Teryaev, Probing Partonic Structure in $\gamma^* \gamma \to \pi \pi$
near Threshold, \prl {\bf 81}, 1782 (1998).

\bibitem{npb555-231}
M.~V.~Polyakov, Hard exclusive electroproduction of two pions and their resonances,
\npb{\bf 555},  231 (1999).

\bibitem{pr49-519}
G.~Breit and E.~Wigner, Capture of slow neutrons, \pr  {\bf 49}, 519 (1936).

\bibitem{prl21-244}
G.~J.~Gounaris and J.~J.~Sakurai, Finite-Width Corrections to the Vector-Meson-Dominance Prediction
for $\rho \to e^+ e^-$, \prl  {\bf 21}, 244 (1968).

\bibitem{zpc62-455}
A.~B.~Clegg and A.~Donnachie, Higher vector meson states produced in electron-positron
annihilation, \zpc {\bf 62}, 455 (1994).

\bibitem{ijmpa13-5443}
M.~K.~Volkov, D.~Ebert, and M.~Nagy, Excited pions, $\rho-$ and $\omega-$ mesons and their decays in a chiral $SU(2)\times
SU(2)$ Lagrangian, \ijmpa{\bf13}, 5443 (1998).

\bibitem{prd88-093002}
N.~N.~Achasov and A.~A.~Kozhevnikov, Pion form factor and reactions $e^+e^- \to \omega \pi^0$
and $e^+e^- \to \pi^+\pi^-\pi^+\pi^-$ at energies up to $2$-$3$ GeV in the many-channel approach,
\prd  {\bf88}, 093002 (2013).


\end{thebibliography}
\end{document}